\documentclass{article}

\usepackage{amssymb,amsfonts,amsmath}
\usepackage{graphics,graphicx}

\begin{document}

\title{Shape- and topology-dependent heat capacity of few-particle systems}
\author{Victor Barsan \\
IFIN-HH, Str. Reactorului no.30, P.O.BOX MG-6, \\
Magurele-Bucuresti, Romania}
\date{April 2, 2012}
\maketitle

\begin{abstract}
Thermal properties of few-fermion ($ n < 5 $) systems are investigated. 
The dependence of the heat capacity on the topology and shape of the cavity
containing the particles is analyzed. 
It is found that the maximum of the heat capacity, occuring at low $ T $, 
discussed recently by Toutounji for a system with $ n = 1 $ fermions, is even 
more visible for $ n = 2 $, but fades away for $ n = 3 $ and $ 4 $. 
For large $ T $, the classical behavior is obtained;
however, when $ T \rightarrow 0 $, the heat capacity tends to zero
exponentially, not linearly, as in macroscopic and even mesoscopic systems.
The physical relevance of these results is discussed.
\end{abstract}

\section{Introduction}

Recent progress in nanophysics has stimulated the interest in studying
particles confined to small volumes or in atypical geometries. 
If we are focused on the theoretical study of thermal properties of such 
systems, the contribution of translational degrees of freedom of the 
constituents to the partition function of the system might be evaluated taking 
into account the discrete character of the particle wave vectors.

To be more specific, let us mention that, for non-interacting many-particle
systems, the translational part of the canonical partition function is
evaluated in the thermodynamic limit, as an integral over the particle
momenta; 
however, for few-particle systems, confined to small cavities, this quantity 
should be obtained as a discrete sum over the quantized values of the particle 
wave vectors. 
Recently, Toutounji \cite{Toutounji} analyzed the thermal properties of an 
electron confined to \ a segment, using the canonical ensemble, showing that 
the heat capacity of this one-particle system has a bump, at "low" 
temperatures, i.e. when $ k_B T $ is much lower than the ground state energy 
(the quotation marks intend to remind that, in this context, the room 
temperature is a "low" temperature). 
Of course, as $ T $ increases, the heat capacity reaches asymptotically the 
value $ k_B / 2 $.
The existence of this bump is smeared out in the thermodynamic limit. 
We can presume that other interesting effects, specific to few-particle systems,
are detectable only taking into account the discrete character of physical
parameters describing such systems. 
A critical discussion of Toutounji's approach, from the perspective of the 
foundations of statistical mechanics, has been done by Lungu \cite{Lungu}.

It is legitimate to wonder about the physical relevance of few--particle
thermodynamics. 
There are several aspects to be discussed: 
(1) is it meaningful to speak about thermal properties of few particle systems,
knowing that thermal physics works essentially with large systems? 
(2) if the response is positive, is the canonical ensemble appropriate for such
investigations? 
(3) there is a real physical interest for the study of thermal behavior of 
few-particle systems?

If the particles are confined to a cavity in thermal equilibrium with a
thermostat, they reach thermal equilibrium not through mutual interaction
(the typical case studied by statistical mechanics), but mainly through
interaction with cavity's walls and with thermal radiation. 
In fact, the calculation of the one-particle partition sum for particles in 
simple external potentials is a common exercise in several books of quantum
mechanics or quantum statistical physics (see for instance \cite{Messiah}, 
\cite{Kubo}).
However, as the thermodynamic relations are obtained, in general, in the
thermodynamic limit, the significance of results referring to few particle
systems, must be considered with caution \cite{Lungu}.

Concerning the second issue, let us recall - invoking again the detailed
analyses done by Lungu \cite{Lungu} - that the grand-canonical system is the 
most appropriate for statistical mechanics calculations, because the quantum
statistical properties of particles are automatically taken into account.
However, we can use the canonical ensemble, if we correctly define the
quantum states and their degeneracies. 
This can be done quite easily for few particles, but becomes a quite cumbersome 
exercise if the number of particles increases. 
It is important to stress the fact that the canonical partition function of a 
quantum few-particle system does not factorize, in terms of one-particle 
partition functions. 
This is why the results obtained for a one-particle system have no relevance 
for several-particle systems.

Finally, the study of few-particle systems is not an academic exercise, as
such systems are available and have very interesting properties, the most
popular being the electrons in quantum dots and atoms in Bose-Einstein
condensates (BEC), see for instance \cite{Okopinska-1}, \cite{Bloch}). 
Investigations in quantum field theory provide another reason for calculating 
the one-particle partition function, for instance of a particle in an 
anharmonic potential \cite{Okopinska-2}. 
Let us also mention that the last paper of Feynman presents a method
for approximating the quantum partition function with a classical one 
\cite{Feynman}.

The starting point of the investigations presented in this paper is the
attempt to reconcile the heat capacity behavior obtained by Toutounji for a
one-electron system - which shows an maximum at low temperature, as already
mentioned - and the well-known aspect of this quantity, for a macroscopic
body, which starts from zero at $ T = 0 $, increases smoothly while $ T $
increases, reaching an horizontal asymptote for $ T \rightarrow \infty $. 
In fact, the same smooth behavior occurs for mesoscopic systems with as few as 
$ n = 14 $ particles, as shown by Lungu \cite{Lungu}. 
So, we are trying to bridge the results obtained by Toutounji, for a system 
with $ n = 1 $ particles, with the results obtained by Lungu, for a system with 
$ n = 14 $ particles. 
We evaluated approximately the canonic partition function, for a system of 2, 3 
and 4 free fermions, taking into account the exclusion principle - so 
considering that the particles are indeed fermions - and found out that the 
low-$ T $ maximum occurring for an one-electron system decreases for systems 
with 3 and 4 particles (even if it is enhanced for a 2-particle system), and the
aspect of the heat capacity plot becomes smoother. 
So, our results bridge the one-particle behavior of the heat capacity, obtained 
by Toutounji, with the mesoscopic behavior, obtained by Lungu, which in fact 
does not differ qualitatively from the macroscopic behavior.

However, there is another "anomaly" which is noticed in the heat capacity of
few-body systems. 
If, for macroscopic systems (see for instance \cite{Kittel}, Ch.6), and even 
for mesoscopic ones \cite{Lungu}, the heat capacity decreases like $ T $ when 
$ T \rightarrow 0 $, in the cases discussed here (1, 2, 3 and 4 particles) it
behaves like $ e^{-1/\tau } / \tau^2 $, where $ \tau $ is a scaled
temperature. 
Such a behavior is compatible with Toutounji's results, even if this issue is 
not explicitly discussed in his paper (see \cite{Toutounji}, Fig. 4).

We also take advantage of the fact that the same - very simple - theoretical
methods, used in Toutounji's paper, can be applied in order to point out the
dependence of the heat capacity on the topology and shape of the system.
After discussing the thermodynamic properties of a particle confined to a
segment, we examine the similar problem, for the same particle confined to a
circle obtained by bending that segment -- so, an 1D body having the same
length, but a different topology. 
As the one-particle energy spectrum are different in the two cases, and as the 
heat capacity is sensitive to the aspect of the energy spectrum, we find that 
the heat capacity of a particle confined to a segment, and on a circle having 
the same length, are different. 
In this way, we obtain a simple example of topology-dependence of the heat 
capacity. 
Then, the same investigations are repeated for systems of 2, 3 and 4 particles, 
having the same confinement as mentioned before.

Such a change of topology can be easily done, for charged particles moving
in a thin toroidal cavity of length $ L $: a bias applied on a small portion 
$ \delta $ of this torus will oblige the particles to move on the rest of the
cavity, topologically equivalent (if the thickness of the cavity tends to
zero) to a segment of length $ L-\delta $. 
With a zero cost in energy, we changed the topology of the body and, implicitly,
its thermal behavior.
Another example is provided by the benzene molecule: if it is broken, the
energy levels are quantized due to the impossibility of electrons to leave
the linear molecule; by contrary, for electrons on the benzene ring, the
quantization results from a cyclic condition \cite{Harrison}. 
The contribution of $ \pi $ electrons to the heat capacity is different for a 
gas of intact benzene molecules, and for one of broken molecules (the ring can 
be broken, for instance, by ultraviolet irradiation).

Another situation under scrutiny refers to few-particle systems confined to
3D cavities: a rectangular (parallelepipedic), cylindrical and spherical
cavity. 
In the first case, the shape of the cavity can be easily modified, at constant 
volume, and the dependence of the heat capacity on the shape of the cavity can 
be easily seen.

The structure of this paper is the following. 
In Section~\ref{sec:2}, the partition sum, internal energy and heat capacity of 
a few-particle system, confined to a segment or to a circle of equal length, 
are obtained and compared.
The values of the heat capacity differ drastically, in the two cases, for low
temperatures, but tend to the same classical value, for high-$ T $. 
In order to avoid irrelevant complications, scaled temperatures have been used. 
All the results are exact, being expressed in terms of the elliptic theta 3
Jacobi function $ \vartheta_3 $ and of its first and second derivatives.
The limiting cases of low- and high-$ T $ are obtained explicitly. 
Section~\ref{sec:3} is devoted to a similar exercise, developed for particles 
confined to 3D cavities, as described in the previous paragraph. 
In the case of the spherical and cylindrical cavities, the results are no more 
exact, as the energy eigenvalues are obtained using an asymptotic approximation 
of the roots of Bessel functions; however, this approximation is quite accurate,
even for the smallest values of the quantum number. 
Also, the summation involved in the evaluation of the partition sums must be 
made over two quantum numbers, running over $ \mathbb{N} $, resulting 
expressions similar to the Jacobi theta functions, but somewhat more 
complicated. 
However, the physical interpretation of results can be done without difficulty. 
In Section~\ref{sec:4}, we present final considerations and conclusions.

\section{A few-fermion system on a segment and on a circle}
 \label{sec:2}

A particle confined to a segment of length $ L $ is described by the
Schrodinger equation for the infinite square well potential:
\begin{equation}
   \label{eq:1}
 V \left( x \right) 
  = \begin{cases}
     0      \; , & x < \frac{L}{2} \\
     \infty \; , & x > \frac{L}{2}
    \end{cases}
\end{equation}
The impenetrability of walls means that:
\begin{equation}
   \label{eq:2}
 \psi \left( - \, \frac{L}{2} \right) 
  = \psi \left( \frac{L}{2} \right) = 0 
\end{equation}
which constraints the wave vector to the values:
\begin{equation}
   \label{eq:3}
 k_{ns} = \frac{n \pi }{L} \; , \quad n = \pm 1 , \pm 2 ,...  
\end{equation}
and consequently produce the energy eigenvalues:
\begin{equation}
   \label{eq:4}
 E_{ns} = \frac{\hbar^2 k_{ns}^2}{2 \, m} 
       = \frac{\hbar^2 \pi^2}{2 \, m \, L^{2}} \, n^{2} 
       = E_{1s} \, n^2   
\end{equation}
where
\begin{equation}
   \label{eq:5}
 E_{1s} = \frac{\hbar^2 \pi^2}{ 2 \, m \, L^2}  
\end{equation}
is the ground state energy of the particle on the segment.

The only effect of including the negative values of $ n $ is a phase factor
multiplying the wave function \cite{Fluegge}, so they can be disregarded.

If the particle moves on a circle of length $ L $, the equation (\ref{eq:2}) is
replaced by a cyclicity condition:
\begin{equation}
   \label{eq:6}
 \psi \left( x \right) = \psi \left( x + L \right) 
\end{equation}
giving:
\begin{equation}
   \label{eq:7}
 k_{nc} = \frac{2 \, n \, \pi }{L} 
\end{equation}
Here and hereafter, the index $ s $ will be used for "segment", and $ c $ -- 
for "circle". The energy levels are:
\begin{equation}
   \label{eq:8}
 E_{nc} = \frac{\hbar^2 \, k_{nc}^2}{2 \, m} 
       = \frac{2 \, \hbar^2 \, \pi^2}{m \, L^2} \, n^2 
       = E_{1c} \, n^2 
\end{equation}
with
\begin{equation}
   \label{eq:9}
 E_{1c} = \frac{2 \, \hbar^2 \, \pi^2}{m \, L^2} 
\end{equation}
is the ground state energy of the particle on the circle. Let us notice
that, as an effect of replacing the rigid-well condition (\ref{eq:2}) with the 
cyclic condition (\ref{eq:5}), we have the relation:
\begin{equation}
    \label{eq:10}
 E_{1c} = 4 \, E_{1s} 
\end{equation}
As the energy of excited states is obtained from the ground state energy by
the same relation, for the segment and for the circle 
(compare Eqs.~(\ref{eq:4}), (\ref{eq:10})),
the energy levels on the circle are 4 times larger that the corresponding
ones on the segment; this fact will have important consequences for the
thermal behavior. The distance between consecutive levels:
\begin{equation}
   \label{eq:11}
 E_{n+1,c} - E_{nc} 
  = \frac{2 \, \hbar^2 \, \pi^2}{m \, L^2} \left( 2 \, n + 1 \right) \; , \quad 
 E_{n+1,s} - E_{ns}
  = \frac{\hbar^2 \, \pi^2}{2 \, m \, L^2} \left( 2 \, n + 1 \right) 
\end{equation}
has of course the same property, reflected directly in the density of states
and heat capacity. 
In fact, the most popular formula expressing the heat capacity in terms of the 
density of states $ g \left( E \right) $ is:
\begin{equation}
   \label{eq:12}
 C = \frac{\partial }{\partial \, T} 
     \int E \; f_{\mathrm{FD}} \left( E,T \right) g \left( E \right) \; 
      \mathrm{d}E  
\end{equation}
It is properly invoked when the energy level are extremely dense, but it is
evident that it shows that, at least at low $ T $, a larger level spacing means
a smaller heat capacity.

As the energy levels are proportional to $ n^2 $, the partition sum for the
particle on a segment, $ Z_s $, and for an identical particle on a circle of
the same length, $ Z_c $, can be expressed as:
\begin{subequations}    \label{eq:13}
 \begin{align}
 Z_s &= \sum_{n=1}^{\infty } \exp \left( - \, \beta \, E_{ns} \right)
      = \sum_{n=1}^{\infty } q_s^{n^2} 
      = Z \left( q_s \right) \; ,   \label{eq:13a} \\
 q_{s} &= \exp \left( - \, \beta \, \frac{\hbar^2 \, \pi^2}{2 \, m \, L^2} 
              \right) 
       = \exp \left( -\beta \, E_{s1} \right) \; ; \label{eq:13b}
 \end{align}
\end{subequations}
\begin{subequations}  \label{eq:14}
 \begin{align}
  Z_c &= \sum_{n=1}^{\infty } \exp \left( - \beta \, E_{nc} \right)
       = q_c^{n^2} 
       = Z \left( q_c \right) \; , \label{eq:14a} \\
   q_c &= \exp \left( - 2 \, \beta \, \frac{\hbar^2 \, \pi^2}{m \, L^2} 
              \right) 
       = \exp \left( -\beta E_{c1} \right) \; . \label{eq:14b}
 \end{align}
\end{subequations}
We shall adopt a more general notation:
\begin{equation}
   \label{eq:15}
 q = \exp \left( - \, \beta \, E_1 \right) 
   = \exp \left( - \, \frac{1}{\tau } \right)
\end{equation}
where we have introduced the scaled temperature $ \tau $:
\begin{equation}
   \label{eq:16}
 \frac{1}{\tau } = \frac{E_1}{k_B T} \; . 
\end{equation}
The sums in Eqs. (\ref{eq:13}) -- (\ref{eq:14}) can be expressed in terms of 
the Jacobi theta 3 function:
\begin{equation}
   \label{eq:17}
 \vartheta_3 \left( 0 , q \right) 
  = 1 + 2 \sum_{k=1}^{\infty } q^{k^2} 
  = 1 + 2 \left( q + q^4 + q^9 + \ldots \right)   
\end{equation}
So, the partition function is:
\begin{equation}
   \label{eq:18}
 Z \left( q \right) 
  = \frac{1}{2} \left[ \vartheta_3 \left( 0 , q \right) - 1 \right]
  = q + q^4 + q^9 + \ldots 
\end{equation}
where $ q $ may be either $ q_s $ or $ q_c $.

Due to Eq.~(\ref{eq:10}), we have the relation:
\begin{equation}
   \label{eq:19}
 q_c = \exp \left( - \, \beta \, E_{1c} \right) 
     = \exp \left( - \, 4 \, \beta \, E_{1s} \right)
     = q_s^4 \; .  
\end{equation}
The thermodynamics can be obtained from Eqs.~(\ref{eq:13}) -- (\ref{eq:14}), 
using the well-known relations:
\begin{equation}
  \label{eq:20}
 U = - \, \frac{\partial \ln Z}{\partial \beta } 
   = k_B T^2 \, \frac{\partial \ln Z}{\partial \, T} 
\end{equation}
\begin{equation}
  \label{eq:21}
 C = \frac{\partial \, U}{\partial \, T} 
   = - \, k_B \, \beta^2 \, \frac{\partial \, U}{\partial \, \beta }
   = - \, k_B \, \beta^2 \, \frac{\partial^2 \ln Z}{\partial \beta^2} \; . 
\end{equation}
As usual, the cases of low and high temperatures are particularly important;
in terms of $ q $, the low-$ T $ limit corresponds to:
\begin{equation}
   \label{eq:22}
 q = \exp \left( - \, \frac{1}{\tau } \right) 
   \sim \exp \left( - \, \frac{1}{0^{+}} \right) 
   \sim \exp \left( - \, \infty \right) 
    = 0   
\end{equation}
and the sum (\ref{eq:17}) is rapidly convergent. 
However, for high-$ T $,
\begin{equation}
   \label{eq:23}
 q = \exp \left( - \, \frac{1}{\tau } \right) 
   \sim \exp \left( - \, \frac{1}{\infty } \right) 
   = e^0 
   = 1 
\end{equation}
and the sum (\ref{eq:17}) converges very slowly. 
In this case, it is convenient to use the following relation, obtained from 
Wolfram - \texttt{EllipticTheta3}(09.03.06.0032.01) \cite{Wolfram}:
\begin{equation}
   \label{eq:24}
 \vartheta_3 \left( 0 , q \right) 
  = \vartheta_3 \left( 0 , \exp \left( - \, \frac{1}{\tau } \right) \right) 
  = \sqrt{\pi \, \tau } \; 
     \vartheta_3 \left( 0 , \exp \left(- \, \pi^2 \tau \right) \right) 
    \; , \qquad  \tau \gg 1 \; . 
\end{equation}
It corresponds to eq. (8) of \cite{Toutounji}.

This means that, for high $ T $, using Eqs.~(\ref{eq:18}) and (\ref{eq:24}), 
the partition sum might be more conveniently written as:
\begin{equation}
  \label{eq:25}
 Z \left( q \right) 
  = \frac{1}{2} \left[ \vartheta_3 \left( 0 , q \right) -1 \right]
  = \frac{1}{2} \left[ \sqrt{\pi \, \tau } \; 
     \vartheta_3 \left( 0 , \exp \left( -\pi^2 \tau \right) \right) - 1 \right]
  \; , \quad q  = \exp \left( - \, \frac{1}{\tau } \right)
\end{equation}
Of course, both Eqs.~(\ref{eq:18}) and (\ref{eq:25}) are exact.

In the limit of low $ T $, and implicitly of small $ q $:
\begin{equation}
   \label{eq:26}
 \ln Z \left( q \right) 
  = \ln q + \allowbreak q^3 - \frac{1}{2} \, q^6 + q^8 + \frac{1}{3} \, q^9 
     - \, q^{11} - \, \frac{1}{4} \, q^{12} + q^{14} + \ldots 
\end{equation}
and from the second equation in (\ref{eq:20}) we get:
\begin{equation}
   \label{eq:27}
 U = E_1 \left[ 1 + 3 \, q^3 - 3 \, q^6 + 8 \, q^8 + \ldots \right] 
\end{equation}
and
\begin{equation}
   \label{eq:28}
 \frac{U}{E_1} \rightarrow 1 \quad \mbox{if } \tau \rightarrow 0 \; . 
\end{equation}
As expected, at $ T = 0 $, the internal energy coincides with the ground state
energy. 
Also, using the first equation in (\ref{eq:21}):
\begin{equation}
    \label{eq:29}
 C = 9 \, k_B \, \frac{1}{\tau ^{2}} \, \exp \left( - \frac{3}{\tau } \right) 
     \left[ 1 - 2 \, \exp \left( - \frac{3}{\tau } \right) 
              + \frac{64}{9} \, \exp \left( - \frac{5}{\tau } \right) 
              + \ldots \right]  
\end{equation}
so
\begin{equation}
    \label{eq:30}
 C \rightarrow 0 \quad \mbox{if } \tau \rightarrow 0 \; , 
\end{equation}
as requested by the third principle of thermodynamics. 
In the $\tau \rightarrow 0 $ limit, the ratio $ C_s / C_c $ is:
\begin{equation}
   \label{eq:31}
 \frac{C_s}{C_c}
  = \frac{1}{16} \, \exp \frac{9 \, E_{1s}}{k_B T} \; . 
\end{equation}
At high $ T $, again with the second equation of (\ref{eq:21}):
\begin{equation}
   \label{eq:32}
 U = \frac{1}{2} \, k_B T 
     + \mathcal{O} \left( \tau \exp \left( - \pi^2 \tau \right) \right)  
\end{equation}
and the heat capacity:
\begin{equation}
   \label{eq:33}
 C = \frac{\mathrm{d} U}{\mathrm{d} T} 
   = \frac{1}{2} \, k_B 
      + \mathcal{O} \left( \tau^2 \exp \left( - \pi^2 \, \tau \right) \right) 
\end{equation}
so the Dulong-Petit law is obtained. 
As a trivial consequence,
\begin{equation}
   \label{eq:34}
 \frac{C_s}{C_c} 
  = 1 + \mathcal{O} \left( \tau^2 \exp \left( - \pi^2 \, \tau \right) \right) 
    \; . 
\end{equation}
Comparing (\ref{eq:31}) with (\ref{eq:34}), it is clear that, at low $ T $, the 
quantum specificity of each system is exponentially dominant, and at high $ T $ 
-- exponentially insignificant.

After getting some physical insight on the physical behavior of the systems,
it is useful to obtain exact expressions for $ U $ and $ C $. 
Let us introduce the following notations:
\begin{align}
 \vartheta_3^{\prime } \left( 0 , q \right) 
  \equiv \frac{\mathrm{d} \vartheta _{3} \left(0 , q \right) }{\mathrm{d} q}
  &= 2 \sum_{k=1}^{\infty } k^2 \, q^{k^2-1}
  = 2 \left( 1 + 4 \, q^3 + 9 \, q^8 + \ldots \right)    \label{eq:35} \\
 \vartheta_3^{\prime \prime } \left( 0 , q \right) 
  = \frac{ \mathrm{d}^2 \vartheta_3 \left( 0 , q \right) }{\mathrm{d} q^2}
  &= 2 \sum_{k=2}^{\infty } k^2 \left( k^2 - 1 \right) q^{k^2-2}
  = 2 \left( 12 \, q^2 + 72 \, q^7 + \ldots \right)   \label{eq:36}
\end{align}
warning about the fact that, traditionally, these symbols are attributed to
the derivatives of $ \vartheta_3 $ with respect to its first argument.
\begin{equation}
   \label{eq:37}
 U = E_1 \, q \, \frac{\vartheta_3^{\prime } \left( 0 , q \right) 
                       }{\vartheta_3 \left( 0 , q \right) - 1}
    \; , \quad \mbox{low } T , \,  q \sim 0 
\end{equation}
\begin{equation}
 U = \sqrt{\pi } \, \frac{E_1}{k_B} \, \tau^2 \, 
     \frac{\frac{1}{2} \, \tau^{-1/2} 
            \vartheta_3 \left( 0 , \mathrm{e}^{ - \pi^2 \tau } \right) 
           - \pi^2 \tau^{1/2} \, \mathrm{e}^{ - \pi^2 \tau } 
              \vartheta_3^{\prime} \left( 0 , \mathrm{e}^{ -\pi^2 \tau } \right) 
          }{\sqrt{\pi \tau } \; 
            \vartheta_3 \left( 0 , \mathrm{e}^{ -\pi^2 \tau } \right) - 1} \; , 
  \; \mbox{high } T , \, q \sim 1  \label{eq:38} 
\end{equation}
Both expressions for $ U $ are of course exact. 
Similarly, for the heat capacity, at low $ T $ it is convenient to use:
\begin{equation}
   \label{eq:39}
 C \left( q \right) 
  = k_B \left( \ln q \right)^2 
    \left[ - \, q^2 \frac{\vartheta_3^{\prime } \left( q \right)^{2}
                          }{\left[ \vartheta_3 \left( q \right) - 1 \right]^2}
           + q \, \frac{\vartheta_3^{\prime } \left( q \right) 
                        }{\left[ \vartheta_3 \left( q \right) - 1 \right] } 
          + q^2 \, \frac{\vartheta_3^{\prime \prime } \left( q \right) 
                         }{\left[ \vartheta_3 \left( q \right) - 1 \right] } 
    \right]  
\end{equation}
and at high $ T $:
\begin{equation}
 C = \sqrt{\pi} \; 
     \frac{\mathrm{d} }{\mathrm{d} \tau } \; 
     \frac{ \frac{1}{2} \, \tau^{3/2} \, 
             \vartheta_3 \left( 0 , \mathrm{e}^{ - \pi^2 \tau } \right) 
           - \, \pi^2 \tau^{5/2} \, \emph{e}^{ - \pi^2 \tau } \, 
              \vartheta_3^{\prime } \left(0 , \mathrm{e}^{ -\pi^2 \tau } \right) 
           }{ \sqrt{\pi \, \tau } \; 
               \vartheta_3 \left( 0 , \mathrm{e}^{ -\pi^2 \tau } \right) -1} 
  \; , \; \mbox{high } T , \, q \sim 1 \; .   \label{eq:40}
\end{equation}
Again, both Eqs.~(\ref{eq:39}) and (\ref{eq:40}) are exact, but the series 
expansion in $ q $ of Eq.~(\ref{eq:34}), and the series expansion in $ \tau $ 
in EWq.~(\ref{eq:40}), are rapidly convergent, at low and high $ T $, 
respectively.

As we could see, the quantities $ T $, $ m $, $ L $ enter through the 
expression:
\begin{equation}
  \label{eq:41}
 \beta \, \frac{\hbar^2 \pi^2 }{m \, L^2} 
  \sim \frac{\beta }{m \, L^2} 
  \sim \frac{1}{m \, L^2 \, T} \, , 
\end{equation}
so the values of $ m $, $ L$ separately are not relevant. 
Replacing the constants with their values and using atomic units:
\begin{equation}
   \label{eq:42}
 2 \, \beta \, \frac{\hbar^2 \pi^2}{m \, L^2}
  =\frac{958.\,\allowbreak 34}{T \, m_a \, L_a^2} \; ,   
\end{equation}
where $ m_a $, $ L_a $ are expressed in amu and {\AA}ngstrom. 
So, the only physical variable entering in the expression (\ref{eq:21}) of the 
heat capacity is the scaled temperature $ \tau $:
\begin{equation}
  \label{eq:43}
 q_s = \exp \left( - \, \beta \, \frac{\hbar^2 \pi^2}{2 \, m \, L^2} \right) 
     = \exp \left( - \, \frac{1}{\tau_s} \right) \; , \quad  
       \tau_s = \frac{T \, m_a \, L_a^2}{240}
\end{equation}
\begin{equation}
   \label{eq:44}
 q_c = \exp \left( - 2 \, \beta \, \frac{\hbar^2 \pi^2}{m \, L^2} \right) 
     = \exp \left( - \, \frac{1}{\tau_c} \right) \; , \quad  
       \tau_c = \frac{T \, m_a \, L_a^2}{958.\,\allowbreak 34} \; . 
\end{equation}
Interesting effects are expected to occur at $ \tau \lesssim 1 $. 
Assuming that $ L_a \sim 10 $~\AA, this means a temperature 
$ T \lesssim 2 / m_a \sim $ 1000~K
 for electrons and about $ 0.1 \ldots 0.01 $~K for atoms.

As the quantity $ q $ takes different values if the particle is confined to a
segment or on a circle, according to Eq.~(\ref{eq:19}), the heat capacity in 
these two cases is different. 
It depends on the topology of the body.

Let us now investigate the situation of a system with more than one fermion,
confined to a segment; in the case of the confinement on a circle, the
results are absolutely similar. 
The energy levels of the system is obtaining filling up the one-particle levels 
and taking into account the degeneracy of states. 
For $ n= 2 , \, 3, \, 4 $ particles, the partition functions are:
\begin{align}
 Z_2 &= \sum_{j=1}^{\infty } \exp \left[ - \frac{2 j^{2}}{\tau } \right]
      + 4 \sum_{j=1}^{\infty } \sum_{k=1}^{\infty } 
          \exp \left[ - \, \frac{j^2 + (j + k )^2}{\tau } \right]
      \; ,    \label{eq:45} \\
 Z_3 &= 2 \sum_{j=1}^{\infty } \sum_{k=1}^{\infty } 
       \exp \left[ - \, \frac{2 j^2 + ( j + k )^2}{\tau } \right] 
      + 2 \sum_{j=1}^{\infty} \sum_{k=1}^{\infty } 
          \exp \left[ - \, \frac{j^2 + 2 ( j + k )^2}{\tau } \right]
            \nonumber \\
     &\;  
      + 8 \sum_{j=1}^{\infty } \sum_{k=1}^{\infty } \sum_{n=1}^{\infty } 
          \exp \left[ - \, \frac{j^2 + ( j + k )^2 
                      + ( j + k + n )^2}{\tau } \right] \; , 
            \label{eq:46} \\
 Z_4 &= \sum_{j=1}^{\infty } \sum_{k=1}^{\infty } 
         \exp \left[ - \, \frac{2 j^2 + 2 ( j + k )^{2}}{\tau } 
              \right]  
        + 4 \sum_{j=1}^{\infty} \sum_{k=1}^{\infty } \sum_{p=1}^{\infty } 
            \exp \left[ - \, \frac{2j^2 + (j + k )^2 
                        + ( j + k + p )^2}{\tau } \right] 
                        \nonumber \\
     &\;
        + 4 \sum_{j=1}^{\infty } \sum_{k=1}^{\infty } \sum_{p=1}^{\infty } 
            \exp \left[ - \, \frac{j^{2} + 2 ( j + k )^2 
                        + ( j + k + p )^2}{\tau } \right] 
                        \nonumber \\
     &\; 
        + 4 \sum_{j=1}^{\infty } \sum_{k=1}^{\infty } \sum_{p=1}^{\infty } 
             \exp \left[ - \, \frac{j^2 + ( j + k )^2 
                         + 2 ( j + k + p )^2}{\tau } \right]
                       \label{eq:47} \\
     &\; 
        + 16 \sum_{j=1}^{\infty } \sum_{k=1}^{\infty } \sum_{p=1}^{\infty} 
             \sum_{q=1}^{\infty } 
             \exp \left[ - \, \frac{j^{2} + ( j + k )^2 
                         + ( j + k + p )^2 
                         + ( j + k + p + q )^2}{\tau } \right] 
                       \nonumber
\end{align}
Clearly, each such expression has the form:
\begin{equation}
   \label{eq:48}
 Z_n = \exp \left( - \, \frac{\mathcal{E}_n}{\tau } \right) 
       \left[ g_{0n} 
             + g_{1n} \, \exp \left(- \, \frac{\varepsilon_{1n}}{\tau } \right)
             + g_{2n} \, \exp \left(- \, \frac{\varepsilon_{2n}}{\tau } \right) 
             + \ldots \right] \; , 
\end{equation}%
where $ \mathcal{E}_n $ is the Fermi energy of the $ n$-particle system, 
$ \varepsilon_{1n} $, $ \varepsilon_{2n} $, \ldots -- the excited states and 
$ g_{jn} $ -- their degeneracies of states. 
They are generalizations of the one-particle partition function:
\begin{equation}
   \label{eq:49}
 Z_1 = \sum_{j=1}^{\infty } \exp \left( - \, \frac{j^2}{\tau } \right) \; . 
\end{equation}
As $ Z_n $ for $ n > 1 $ cannot be expressed in terms of a finite number of
known functions, for the evaluation of the heat capacity, the series in
Eqs.~(\ref{eq:45}) -- (\ref{eq:47}) will be cut at some value of 
$ l $, $ k $, etc. 
So, it is useful to have an idea of errors introduced in this way.

Let $ Z_n \left( \max = J \right) $ denote the finite sum obtained from $Z_n $,
if we keep only the terms of the sums with summation indices not larger
than $ J $. 
The only case when the finite sum can be compared with the exact value is 
$ n = 1 $. 
We find that $ \ln Z_1 \left( \max =100 \right) $ approximates $ \ln Z_1 $ with 
an error less than $ 10^{-15} $ for $ \tau <10 $
and $ \ln Z_1 \left( \max =10 \right) $ approximates $ \ln Z_1 $ with an
error less than $ 10^{-13} $ for $ \tau < 4 $ and less than $ 10^{-7} $ for 
$ \tau < 8 $. 
So, we can presume that, for a qualitative description of the heat capacity, it 
will be acceptable to use the sums $ Z_n \left( \max = 10 \right) $. 
However, for $ n = 1 $, we have used $ Z_n \left( \max = 100 \right) $; 
we avoided to work with the exact partition sum, as the exact calculation of
the heat capacity involves the derivative of $ \vartheta_3 $ with respect
with its second argument, which is a very complicated function (not to be
confused with EllipticTheta Prime 3) and is not very appropriate to be used
in Mathematica calculations.

It is convenient to use the scaled temperature $ \tau $ instead of the
physical one $ T $, and to define an effective heat capacity per particle:
\begin{equation}
   \label{eq:50}
 \widetilde{c}_n 
  = \frac{1}{n} \; \frac{\partial \, \widetilde{U}_{n}}{\partial \, \tau} 
  = \frac{1}{n} \; \frac{\partial }{\partial \tau} 
     \left( \tau^2 \, \frac{\partial \ln Z_{n}}{\partial \, \tau }\right) \; . 
\end{equation}
As the partition sums $ Z_{n} $, $ n > 1 $, cannot be expressed in terms of 
$ \vartheta_3 $, or other special functions of mathematical physics, the
internal energy, the heat capacity etc. cannot be written in compact form.
However, due to the fact that $ Z_{n} $, $ n > 1 $, have similar behaviors near 
$ \tau = 0 $, (compare Eqs.~(\ref{eq:48}), (\ref{eq:49}), we still shall have:
\begin{equation}
  \label{eq:51}
 \widetilde{c}_{n}
  \sim \frac{\exp \left( - \, \frac{3}{\tau } \right) }{\tau^2} \; , 
\end{equation}
which differs from the power low $ T$-dependence which occurs for macroscopic
and even microscopic systems.

\begin{figure}[h]
  \centering
  \includegraphics[width=1\textwidth]{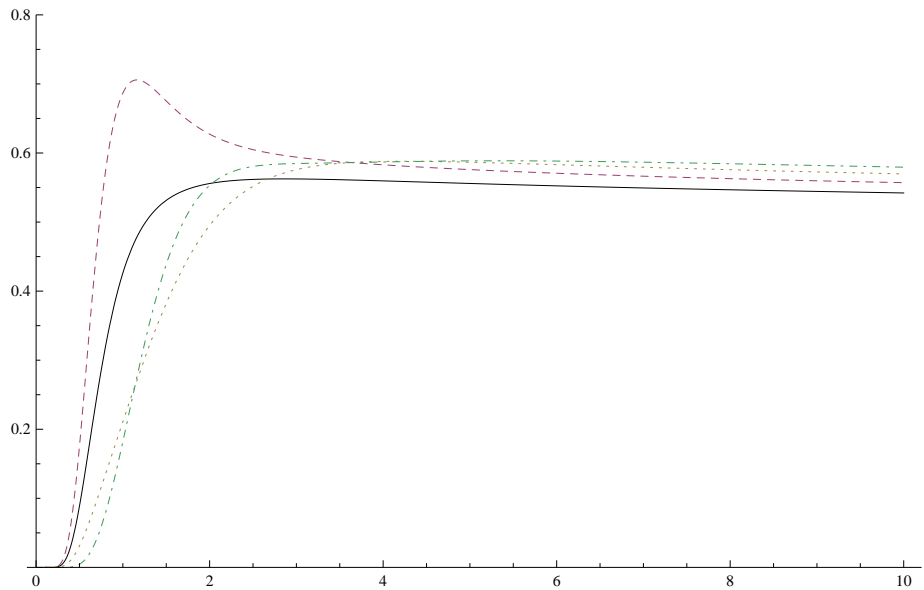}
  \caption{The heat capacity per particle for a system of $ n $ fermions; 
    $ n = 1 $ : solid line; $ n = 2 $ : dashed; $ n = 3 $ : dotted; $ n = 4 $ : 
    dot - dashed.}
\label{fig:1}
\end{figure}

The temperature dependence of the heat capacity, for systems with 
$ n = 1 $, $ 2 $, $ 3 $ and $ 4 $, is represented in Figure~\ref{fig:1}. 
The heat capacity per particle is plotted with solid, dashed, dotted, 
dot-dashed line, for $ n= 1 $, $ 2 $, $ 3 $, $ 4 $ respectively. 
The maximum value of $ \widetilde{c}_n $ for $ n = 1 $, $ 2 $, $ 3 $, $ 4 $
occur at the following values of $ \tau $:
\begin{equation}
   \label{eq:52}
 2.84 \, , \; 1.16 \, , \; 4.25 \, , \; 5.2 \; .
\end{equation}
Their relative heights, defined as:
\begin{equation}
   \label{eq:53}
 \frac{\widetilde{c}_{n,\max } - \widetilde{c}_n \left( \tau = 10 \right) 
       }{\widetilde{c}_n \left( \tau = 10 \right) } 
\end{equation}
are:
\begin{equation}
    \label{eq:54}
  0.0376 \, , \; 0.2671 \, , \; 0.0321 \, , \; 0.0157 \; . 
\end{equation}
So, the effect manifests strongly at $ n = 2 $, but fades out for larger values
of $ n$. 
For large temperature, they show the same tendency as mesoscopic and
macroscopic systems, i.e. tends to a horizontal asymptote, satisfying
Dulong-Petit law.

Two comments concerning the plots in Figure~\ref{fig:1} should be done. 
First, it is quite unexpected that the maximum of the heat capacity is more 
visible for $ n = 2 $ than for $ n = 1 $. 
Second, we must be aware of the fact that $ Z_n \left( \max = 10 \right) $ are 
good approximations for small $ \tau $, so clearly more credible for 
$ \tau \lesssim 1 $ than for $ \tau \sim 10 $.

Another generalization of our initial one-fermion problem can be obtained
considering, instead of a particle confined to a segment, the same particle
confined to an elongated rectangle with edges $ a \, , \, b \; ; \, a \ll b $. 
The quantization along the axes $ Ox \, , \, Oy $ gives:
\begin{equation}
    \label{eq:55}
 k_{x} = \frac{\pi }{a} \; , \quad 
 k_{y} = \frac{\pi }{b} \; , 
\end{equation}
with the ground state one-particle energy levels:
\begin{equation}
    \label{eq:56}
 E_{1sx} = \frac{\hbar^2 \, \pi^2}{2 \, m \, a^2} \; , \quad 
 E_{1sy} = \frac{\hbar^2 \, \pi^2}{2 \, m \, b^2} \; . 
\end{equation}
The partition sum factorizes:
\begin{equation}
    \label{eq:57}
 Z = Z \left( q_{sx} \right) \; Z \left( q_{sy} \right) \; , \quad  
 q_{x} = \exp \left( -\beta E_{1sx} \right) \; , \quad 
 q_{y} = \exp \left( -\beta E_{1sy} \right) \; . 
\end{equation}
or:
\begin{equation}
    \label{eq:58}
 Z = \frac{1}{4} \left[ \vartheta_3 \left( q_{sx} \right) - 1 \right] \, 
                 \left[ \vartheta_3 \left( q_{sy} \right) - 1 \right] \; . 
\end{equation}
The heat capacity is a sum of contributions from each direction:
\begin{equation}
  \label{eq:59}
 C \left( q_{sx} , q_{sy}\right) 
  = C \left( q_{sx} \right) + C \left( q_{sy} \right) \; ,  
\end{equation}
where $ C \left( q_{sx} \right) $ is given by Eq~(\ref{eq:39}). 
If we keep fixed one of the long edges of the rectangle, and curb the rectangle,
superposing the other long edge on the fixed one, we get an empty cylinder; 
if we bend the cylinder, superposing its ends, we get an empty torus. 
For a particle moving on its surface, the quantization of its wavevector is 
similar to Eq.~(\ref{eq:7}), and its thermodynamics can be obtained from the 
previous relations, with the index replacement $ s \rightarrow c $. 
As $ q_c = q_s^4 $, we have again a situation when the heat capacity depends on 
the topology of the body. 
In principle, such a situation can be encountered in the case of a rectangle of
graphene, and of a toroidal nanotube. 
The translational electronic contribution to the heat capacity for these two 
bodies will be different. 
If the heat capacity of such bodies can be measured with a precision of about 
$ 10^{-4} $, this effect can be observed experimentally.

\bigskip

\section{Heat capacity of a few-fermion system confined to various 
            cavities}
\label{sec:3}

Let us firstly consider a fermion in a cavity having the form of a
rectangular prism, with square basis of area $ a \times a $ and height $ b $. 
We shall assume that the volume $ V = a^2 b $ is constant, and the cavity is
described by the parameters $ a $, $ b = V/a^2 $. 
The movement of a particle inside the cavity is quantized according to the 
relations:
\begin{equation}
  \label{eq:60}
 k_x = k_y = \frac{\pi }{a} \; , \quad 
 k_z = \frac{\pi }{b} = \frac{\pi a^2}{V} \; .  
\end{equation}
Putting:
\begin{equation}
   \label{eq:61}
 \frac{a^3}{V} = \varepsilon \; , 
\end{equation}%
the energy levels corresponding to each axis are:
\begin{align}
 E_{1x} = E_{1y} = \frac{\hbar^2 \, \pi^2}{2 \, m \, a^2} \; , \quad 
 E_{1z} = E_{1x} \; \varepsilon^2 \; ,  \label{eq:62} \\
 E_{nx} = E_{ny} = E_{1s} \, n^2 \; , \quad  
 E_{nz} = \varepsilon^2 \, E_{1s} \, n^2 \; .   \label{63}
\end{align}
If $ \varepsilon \ll 1 $, the levels corresponding to the $ Oz $ axis are very
dense, so they are give the dominant contribution to the heat capacity, but
if $ \varepsilon \gg 1 $, they are very rare, and their contribution is
negligible. 
The partition sum factorizes:
\begin{equation}
   \label{eq:64}
 Z = Z \left( q_x \right)^2 \; Z \left( q_z \right) \; ,
 \begin{cases}
  q_x = \exp \left( - \, \beta \, E_{1x} \right) = q_y \\
  q_z = \exp \left( - \, \beta \, \varepsilon \, E_{1z} \right) 
       =\left( q_x \right)^{\varepsilon }
 \end{cases}
\end{equation}
or:
\begin{equation}
  \label{eq:65}
 Z = \frac{1}{8} \left[ \vartheta_{3s} \left( q_x \right) -1 \right]^2 
     \left[ \vartheta_3 \left( \left( q_x \right)^{\varepsilon } \right) -1 
     \right] \; .  
\end{equation}
Due to this property, the heat capacity is a sum of contributions from each
direction:
\begin{equation}
   \label{eq:66}
 C \left( q , \varepsilon \right) 
  = - \, k_B \beta^2 \left[ 2 \, 
      \frac{\partial^2 \ln Z \left( q_x \right) }{\partial \beta^2}
     + \frac{\partial^2\ln Z \left( q_x^{\varepsilon } \right) 
             }{\partial \beta^2 } \right] 
  = 2 \, C \left( q_x \right) + C \left( q_x^{\varepsilon } \right) \; . 
\end{equation}
Let us put, similar to previous cases:
\begin{equation}
   \label{eq:67}
 \frac{1}{\tau_x} = \frac{E_{1x} }{k_B T} \, , \qquad 
 \frac{1}{\tau_z} = \frac{E_{1z} }{k_B T} 
                   = \frac{\varepsilon^2}{k_BT} \; . 
\end{equation}
In the low-$ T $ limit, taking into account Eq.~(\ref{eq:29}), we have the 
following limiting cases:

If $ \varepsilon \ll 1 $,
\begin{equation}
   \label{eq:68}
 \frac{1}{\tau_z} \ll \frac{1}{\tau_x} \; , \qquad 
 C \left( q_s^{\varepsilon} \right) = C_z \gg C_x \; .  
\end{equation}
If $ \varepsilon \gg 1 $,
\begin{equation}
   \label{eq:69}
 \frac{1}{\tau_z} \gg \frac{1}{\tau_x} \; , \qquad 
 C \left( q_s^{\varepsilon} \right) = C_z \ll C_x \; . 
\end{equation}
In the high-$ T $ limit, Eqs.~(\ref{eq:66}) and (\ref{eq:33}) give:
\begin{equation}
   \label{eq:70}
 C \left( q , \varepsilon \right) 
  = \frac{3}{2} \, k_B 
    + \mathcal{O} \left( \tau^2 \, \exp \left( - \, pi^2 \, \tau \right) \right)
       \; ,  
\end{equation}%
so Dulong-Petit law is obtained, in its 3D form.

If the prism is very long, $ a \ll b $ or $ \varepsilon \ll 1 $, it can be bent
until its opposite faces superpose each other, forming an empty ring; 
the quantization corresponding to the $ Oz $ axis is dominant, and the behavior 
of the system tends to that of a particle on a circle.

The case of a particle in a spherical, cylindrical or toroidal cavity can be
treated similarly to the 1D problems discussed in Section 2. 
The situation is of course more complicated, for two reasons. 
First, the energy eigenvalues can be obtained only approximately, from the 
asymptotic values of zeros of Bessel functions (of half-integer order, for a 
sphere, and of integer order, for cylinder and torus). 
Second, the summation involves three quantum numbers; 
the azimutal one gives a trivial contribution, but the other two run over all 
integers, from 1 to $ \infty $. 
The final result is that the partition sum is not expressed any more in terms 
of $ \vartheta_3 $, but in terms of more general functions:
\begin{equation*}
 \vartheta \left( q ; \alpha , \beta \right) 
  = \sum n^{\alpha } \, q^{\left( n+\beta \right)^2}
\end{equation*}
with $ \alpha =0 , 1 , 2 $ and $ \beta = 0 , \ 1/2 $. 
We omit the detailed presentation of these functions, which are quite simple, 
but cumbersome and of small physical interest. 
The heat capacity of a particle confined to a sphere or a cylinder is quite 
similar to that presented in Figure~\ref{fig:1}.

\section{Conclusions}
\label{sec:4}

The starting point of the investigations presented in this paper is the fact
that, for few-particle systems confined to a small volume, their thermal
behavior must be evaluated using a partition sum calculated discretely. In
order to be able to treat the problem exactly, or at least in controllable
approximations, only the free particles (in fact fermions) without internal
structure have been considered. In this case, only the translational degrees
of freedom contribute to the partition function. According to the remark
just made, the partition function of a small-volume system should not be
evaluated as an integral over momenta, as usually done for large systems,
but as a sum over the quantized values of the particles wave vectors.

As recently discussed by Toutounji \cite{Toutounji}, if we evaluate in this way 
the canonical partition sum for an one-fermion system, confined to a segment,
and compute the heat capacity using the standard thermodynamic formulas, we
find that this quantity shows a maximum at low temperature. Even if, with
increasing $T$, the heat capacity tends to a horizontal asymptote, its low-$ T $
behavior differs qualitatively of that of a macroscopic body, which
increases smoothly from zero, at $ T = 0 $, to a constant value, at large
temperatures, according to \ the Dulong - Petit law. In fact, the same
macroscopic behavior occurs, at least qualitatively, for mesoscopic systems
too, as shown recently by Lungu, who considered systems of 14, 76 and 820
particles \cite{Lungu}.

In this respect, our paper reveal two new aspects. First, it shows that the
heat capacity falls at $ T = 0 $ according to an exponential law, merely than a
power law, as in mesocsopic and macroscopic systems. Second, the maximum of
the heat capacity, discussed by Toutounji for a system with \thinspace $ n = 1 $
particles, is even stronger for $ n = 2 $, but fades out for $ n = 3 $ and 
$ n = 4 $.

Using the same theoretical methods, we calculated the heat capacity of
few-body systems having the same local geometric properties, but different
topology - for instance, a segment and a circle with identical length, or a
cylinder and a torus obtained by bending this cylinder - and we found that
the heat capacity depends on the topology of the body. Also, few-particle
systems confined to cavities having the same volume, but different
geometries are different. Even if these conclusions might be surprising at
first sight, they are simple effects of the sensitivity of the thermal
properties of few-particle systems to the specificities of energy spectrum,
at low T, and do not contradict the well known results valid for macroscopic
bodies, obtained theoretically after the thermodynamic limit $\left(
N\rightarrow \infty ,\ V\rightarrow \infty \right) $ is taken.

In the same time, these results are simple and pedagogical illustrations for
the peculiar behavior of nanoscopic systems, as compared to macroscopic, and
even with mesoscopic ones. These effects - for instance topology-dependent
heat capacity - could be observed experimentally, in systems composed of
benzene molecules or nanotubes.

\section*{Acknowledgements}

The author is indebted to Prof. R.P. Lungu, for illuminating discussions,
and to PhD student R. Dragomir, for useful remarks. The financial support of
the Romanian National Authority for Scientific Research (ANCS),
CNCS-UEFISCDI, project IDEI number 953 / 2008, of the ANCS project PN 09 37
01 06 and of JINR Dubna - IFIN-HH Magurele-Bucharest project no.
01-3-1072-2009/2013 are kindly acknowledged.

\end{document}